\documentclass{iaus}
\usepackage{graphicx}

\title[The $M_\star$--$Z_\star$ relation in SDSS galaxies] 
{What stellar populations can tell us about the evolution of the
mass--metallicity relation in SDSS galaxies}

\author[Vale Asari et al.]   
{N. Vale Asari$^{1,2}$,
  G. Stasi\'nska$^{2}$,
  R. Cid Fernandes$^{1}$,
  J. M. Gomes$^{1,3}$,
  M. Schlickmann$^{1}$,
  A. Mateus$^{4}$,
  W. Schoenell$^{1}$}

\affiliation{$^{1}$Dpto.\ de F\'{\i}sica - CFM - Universidade Federal de
  Santa Catarina, Florian\'opolis, SC, Brazil\\
  $^{2}$LUTH, Observatoire de Paris, CNRS, Universit\'e Paris
  Diderot; Place Jules Janssen 92190 Meudon, France\\
  $^{3}$GEPI, Observatoire de Paris, CNRS, Universit\'e Paris
  Diderot; Place Jules Janssen 92190 Meudon, France\\
  $^{4}$IAG, Universidade de S\~ao Paulo, S\~ao Paulo, SP,
  Brazil}

\begin{document}

\maketitle

\begin{abstract}
During the last three decades, many papers have reported the existence
of a luminosity--metallicity or mass--metallicity ($M$--$Z$) relation
for all kinds of galaxies: The more massive galaxies are also the ones
with more metal-rich interstellar medium.  We have obtained the
mass-metallicity relation at different lookback times for the same set
of galaxies from the Sloan Digital Sky Survey (SDSS), using the
stellar metallicities estimated with our spectral synthesis code {\sc
starlight}. Using stellar metallicities has several advantages: We are
free of the biases that affect the calibration of nebular
metallicities; we can include in our study objects for which the
nebular metallicity cannot be measured, such as AGN hosts and passive
galaxies; we can probe metallicities at different epochs of a galaxy
evolution.

We have found that the $M$--$Z$ relation steepens and spans a wider
range in both mass and metallicity at higher redshifts for SDSS
galaxies. We also have modeled the time evolution of stellar
metallicity with a closed-box chemical evolution model, for galaxies
of different types and masses. Our results suggest that the $M$--$Z$
relation for galaxies with present-day stellar masses down to $10^{10}
M_\odot$ is mainly driven by the star formation history and not by
inflows or outflows.

\end{abstract}

A full version of this study is presented in \cite{ValeAsari_etal_2009}.

\begin{figure}[ht]
  \begin{center}
   \includegraphics[width=\textwidth, bb=40 170 592 310]{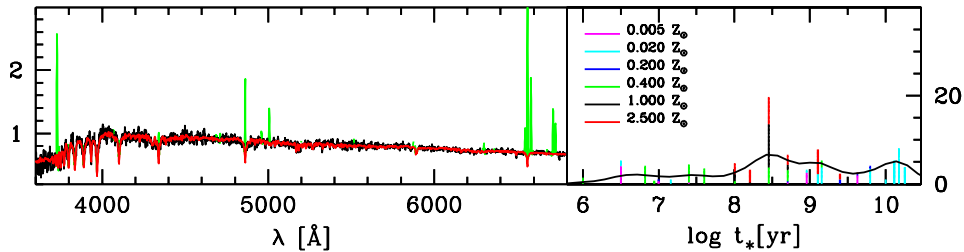}
  \end{center}
  \caption{We have processed 573141 SDSS galaxies with our
    pixel-by-pixel {\sc starlight} algorithm, which fits a
    galaxy spectrum with a sum of simple stellar populations (SSPs) of
    different ages and metallicities; Left panel shows an example of
    observed (black+green) and model (red) spectra. Emission lines
    (green) are masked out. Right panel shows the light fraction (in
    percents) associated with the SSPs.  See
    \cite{CidFernandes_etal_2005} for more details about the method.}
\label{fig:fits}
\end{figure}

\begin{figure}[ht]
  \begin{center}
    \includegraphics[width=0.4\textwidth, bb=40 535 405 695]{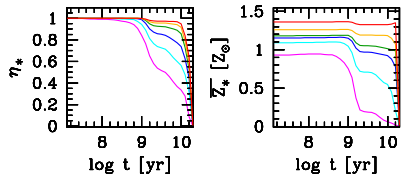}
  \end{center}
  \caption{From the ages and metallicities of the SSPs in a galaxy we
    recover histories of stellar mass and stellar metallicity. Our
    sample has been divided into six present-day stellar mass bins
    centered in $\log M_\star/M_\odot = 10.0$ (A--magenta), 10.3
    (B--cyan), 10.6 (C--blue), 10.9 (D--green), 11.2 (E--yellow) and
    11.5 (F--red), each one 0.30 dex wide.  Panels show the cumulative
    stellar mass history ($\eta_\star(t) \equiv M_\star(t) /
    M_\star(t=0)$, left) and the mean stellar metallicity
    ($\overline{Z_\star}(t)$, right) as a function of lookback time
    $t$ for bins A--F.  We find that the {\em more massive} a galaxy
    is today, the {\em faster} it has formed stars and produced
    metals.}
\label{fig:histories}
\end{figure}

\begin{figure}[ht]
  \begin{center}
   \includegraphics[width=0.7\textwidth, bb=40 535 592 695]{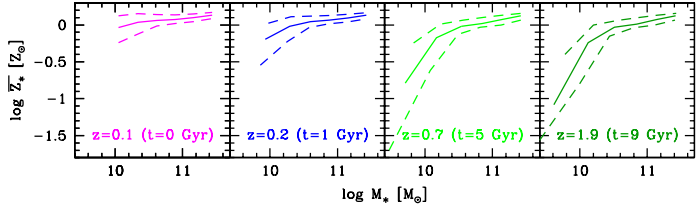}
  \end{center}
  \caption{Given $\eta_\star(t)$ and $\overline{Z_\star}(t)$ for a set
    of galaxies, another way to study their evolution is to look at a
    snapshot of the $M_\star$--$Z_\star$ relation for a given lookback
    time $t$.  Solid lines mark the median and dashed lines the
    quartiles of the distribution.  As lookback time increases, the
    $M_\star$--$Z_\star$ relation steepens and covers a larger range
    of values.  The novelty is that we are looking at the {\em same
      set} of galaxies in each lookback time. That was only possible
    because we have derived {\em stellar} instead of {\em nebular}
    metallicities.}
\label{fig:redshift}
\end{figure}

\begin{figure}[ht]
  \begin{center}
   \includegraphics[width=0.72\textwidth, bb=30 455 550 710]{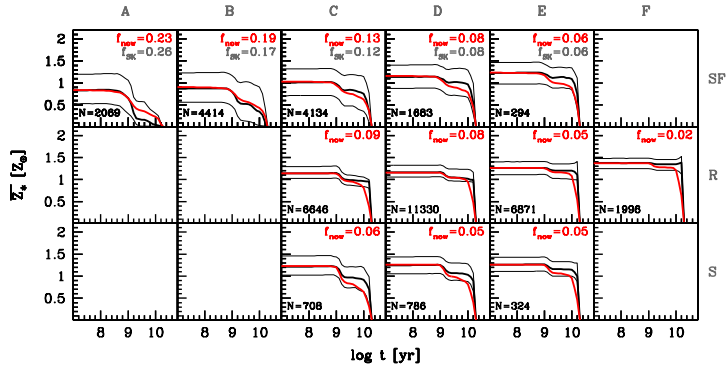}
  \end{center}
  \caption{We show a simple closed-box chemical evolution model for
    mass-bins A--F and by galaxy type: star-forming (SF),
    retired/passive (R) and Seyfert (S).  Each panel shows the median
    and quartiles of the evolution of $Z_\star$ as found by {\sc
      starlight} (black lines) and as obtained with the simple
    closed-box model (red lines).  The value of the present-day gas
    fraction ($f_{\rm now}$) needed to reproduce the median
    present-day $Z_\star$ is indicated at the top right of each
    panel. For the SF sample, we also indicate the median of $f_{\rm
      now}$ as measured by the Schmidt-Kennicutt law ($f_{\rm SK}$).
    This suggests that the $M_\star$--$Z_\star$ relation for
    galaxies with present-day stellar masses down to $10^{10} M_\odot$
    is mainly driven by the star formation history.  }
\label{fig:closedbox}
\end{figure}

\end{document}